\documentclass[twocolumn,aps,pra,groupedaddress,showpacs]{revtex4}
\usepackage{epsfig,amssymb}
\def\comment#1{}\def\labell#1{\label{#1}}
\begin{document}
\title{Extended phase-matching conditions for improved entanglement  
generation}
\author{Vittorio Giovannetti, Lorenzo Maccone, Jeffrey  H. Shapiro,
and Franco N. C. Wong}
\affiliation{\mbox{Massachusetts Institute of Technology, Research
Laboratory of Electronics,} \\Cambridge, Massachusetts 02139.}
\date{\today}

\begin{abstract}
Extended phase-matching conditions for spontaneous parametric
down-conversion are examined. By augmenting the conventional
phase-matching conditions, they permit the creation of a class of
frequency-entangled states that generalizes the usual twin-beam
biphoton state. An experimental characterization of these states is
possible through interferometric coincidence counting.
\end{abstract}
\pacs{42.50.Dv, 03.65.Ud, 03.67.-a, 42.25.Hz}
\maketitle

\section{Introduction}\labell{s:intro}
Entanglement is the cornerstone of quantum information technology. Its
wide-ranging applications include tests of the foundations of quantum
mechanics, implementations of quantum algorithms, cryptographic
communication procedures, measurement-accuracy enhancements, {\it
etc.} Arguably, the most important source of entanglement is
spontaneous parametric down-conversion (SPDC) in nonlinear optical
crystals. In the down-conversion process, a photon from an intense
pump beam is absorbed in a nonlinear crystal and two different photons
(conventionally called signal and idler) are created. Energy
conservation dictates that the frequencies of the signal and idler
photons sum to that of the pump photon. Frequency entanglement arises
from quantum superposition of the various ways in which this energy
constraint is fulfilled by signal and idler photons belonging to a
certain frequency interval. This interval is governed by
phase-matching within the nonlinear crystal, a condition which is
dependent on the crystal's length and the refractive indices along its
principal optical axes {\cite{mandel}}.

In this paper we show how to obtain new kinds of
frequency-entanglement by generalizing the conventionally used
phase-matching condition. The resulting biphoton states, which were
previously obtained by Erdmann, Branning, Grice and Walmsley in
{\cite{walms}}, can be characterized using two interferometers based
on the Hong-Ou-Mandel and on the Mach-Zehnder setups, as described
below. As will be shown, the extended conditions can be fulfilled with
available crystals. They allow the use of long nonlinear crystals that
improve the generation efficiency. Furthermore, they allow the use of
short-duration pump pulses that increase the experimental signal-idler
generation rate compared to that of continuous-wave (cw)
pumping. These two advantages are not readily available with the
conventional phase-matching condition: on one hand the use of long
type II phase-matched crystals narrows the spectrum over which the
down-converted photons are entangled, and, on the other hand, the use
of pulsed pumping reduces the entanglement because the
indistinguishability of the down-converted photons is adversely
affected {\cite{keller,wal,ata}}.

The paper is organized as follows. In Sect.~{\ref{s:conv}} we
introduce our notation and give a brief derivation of the conventional
phase-matching condition. In Sect.~{\ref{s:ext}} we derive the
extended phase-matching conditions and analyze their role in obtaining
frequency entanglement. The following Sect.~{\ref{s:pol}} shows how
one can obtain polarization entanglement starting from the states
found in Sect.~\ref{s:ext}. In Sect.~{\ref{s:car}} two experimental
setups that can be used to characterize the frequency entanglement are
described and analyzed. Appendix~\ref{s:app} closes the paper with the
detailed visibility derivation for the interference experiments
discussed in Sect.~\ref{s:car}.

\section{Biphoton state via parametric
down-conversion}\labell{s:conv} In this paper we employ continuous
Fock space formalism, in which the photon annihilation operator
$a_j(\omega)$ that destroys a photon of frequency $\omega$ in the
$j$th mode obeys the commutation relation
\begin{eqnarray} \left[a_j(\omega),a_k^\dag(\omega')\right]
=\delta(\omega-\omega')\;\delta_{jk}
\;\labell{commutaz}.
\end{eqnarray}
To first order in the nonlinear interaction coupling $\chi^{(2)}$, the
state of the system at the output of a compensated down-conversion
crystal is the biphoton state {\cite{walms,mandel}}
\begin{eqnarray}
|\Psi\rangle &=&
\int\! \frac{\displaystyle d\omega_s}{\displaystyle 2\pi}
\int\! \frac{\displaystyle d\omega_i}{\displaystyle 2\pi}\;
A(\omega_s,\omega_i)\;a^\dag_s(\omega_s)\;a^\dag_i(\omega_i)\;
|0\rangle,
\labell{stato}
\end{eqnarray}
where $a_s(\omega)$ and $a_i(\omega)$ are the annihilation operators
for signal $s$ and idler $i$ modes, respectively, and $|0\rangle$ is
the vacuum state. In Eq.~(\ref{stato}) the biphoton spectral amplitude
$A(\omega_s,\omega_i)$ determines the frequency spectrum of the
biphoton state: integrating its squared modulus over the frequency of
one of the two modes yields the fluorescence spectrum of the other. If
we neglect inconsequential normalization constants and assume colinear
plane-wave propagation, $A(\omega_s,\omega_i)$ can be written as
(employing the notation from {\cite{noi}})
\begin{eqnarray} A(\omega_s,\omega_i)=\alpha(\omega_s,\omega_i)\;
\Phi_L(\omega_s,\omega_i)
\;\labell{defa},
\end{eqnarray}
where
\begin{eqnarray}
\alpha(\omega_s,\omega_i)&\equiv&\frac{\sqrt{\omega_s\omega_i}}
{n_s(\omega_s)n_i(\omega_i)}\,
{\cal E}_p(\omega_s+\omega_i)\labell{alpha}\;,\\
\Phi_L(\omega_s,\omega_i) &\equiv&
\frac{\sin\left(\Delta k(\omega_s,\omega_i)L/2\right)}
{\Delta k(\omega_s,\omega_i)/2}
\labell{phi}\;.
\end{eqnarray}
The quantity $\alpha(\omega_s,\omega_i)$ depends on the refractive
indices, $n_s$ and $n_i$, of the signal and idler modes in the
nonlinear crystal, and on the pump fluorescence spectrum, $|{\cal
E}_p(\omega)|^2$, which is centered at $\omega_p$ and has a bandwidth
$\Omega_p$. The refractive indices, together with the term
$\sqrt{\omega_s\omega_i}$, can in general be considered as constants
over the effective spectrum, so that $\alpha(\omega_s,\omega_i)$ is
approximated by a function of the sum of the signal and idler photon
frequencies, $\omega_s+\omega_i$.  On the other hand, the quantity
$\Phi_L(\omega_s,\omega_i)$ is the phase-matching function, which
depends on the nonlinear crystal length $L$ and on the phase mismatch
\begin{eqnarray}
\Delta k(\omega_s,\omega_i)\equiv
k_p(\omega_s+\omega_i)-k_s(\omega_s)-k_i(\omega_i)\;\labell{deltak},
\end{eqnarray}
where $k_{p,s,i}(\omega)\equiv\omega n_{p,s,i}(\omega)/c$ denote the
wave numbers of the pump, signal and idler photons,
respectively. Notice that $k_p$, $k_s$ and $k_i$ are not all
independent in the birefringent crystals that are typically used for
SPDC where $k_s(\omega)=k_i(\omega)$ for type I phase-matching, and
$k_p(\omega)=k_s(\omega)$ (or $k_p(\omega)=k_i(\omega)$) for type II
phase-matching.

For frequency entanglement generation, the SPDC crystal is usually
operated at frequency degeneracy, in which the signal and idler
spectra are centered around half the mean frequency of the pump
$\omega_p$ {\cite{keller,wal,ata}}. This is enforced by imposing the
conventional phase-matching condition that requires the crystal's
refractive indices to satisfy
\begin{eqnarray}
n_p(\omega_p)
&=&\frac{n_s(\omega_p/2)+n_i(\omega_p/2)}2\;.
\labell{ordine0}
\end{eqnarray}
Employing Eq.~(\ref{ordine0}) and approximating the function $\Delta
k$ of Eq.~(\ref{deltak}) with a first-order Taylor expansion in
$\omega_s$ and $\omega_i$ around $\omega_p/2$, we obtain
\begin{eqnarray}
\Delta k(\omega_s,\omega_i)=(\omega_s-\omega_p/2)\gamma_s+
(\omega_i-\omega_p/2)\gamma_i
\;\labell{deltaktaylor},
\end{eqnarray}
where $\gamma_s\equiv k'_p(\omega_p)-k'_s(\omega_p/2)$ and
$\gamma_i\equiv k'_p(\omega_p)-k'_i(\omega_p/2)$. With this result,
the biphoton state $|\Psi\rangle$ of Eq.~(\ref{stato}) can be written
as
\begin{widetext}
\begin{eqnarray} |\Psi_{{ pm}}\rangle &=&
\int\! \frac{\displaystyle d\tilde\omega_s}{\displaystyle 2\pi}
\int\! \frac{\displaystyle d\tilde\omega_i}{\displaystyle 2\pi}\;
\alpha(\tilde\omega_s+\tilde\omega_i+\omega_p)
\;\frac{\sin\left[(\gamma_s\tilde\omega_s+\gamma_i\tilde\omega_i)L/2\right]}
{\left(\gamma_s\tilde\omega_s+\gamma_i\tilde\omega_i\right)/2}
\;|\omega_p/2+\tilde\omega_s\rangle_s\;|\omega_p/2+\tilde\omega_i\rangle_i,
\;\labell{pmstato}
\end{eqnarray}
\end{widetext}
where $\tilde\omega_s=\omega_s-\omega_p/2$ and
$\tilde\omega_i=\omega_i-\omega_p/2$ represent the signal and idler
detunings from degeneracy, and $|\omega\rangle\equiv
a^\dag(\omega)|0\rangle$.

If the crystal is pumped with monochromatic light $\Omega_p\to 0$,
then $\alpha(\omega)\to\delta(\omega-\omega_p)$ and the state
$|\Psi_{pm}\rangle$ becomes \begin{eqnarray}|{\rm
TB}\rangle\equiv\int
\frac{\displaystyle d\tilde\omega}{\displaystyle
2\pi}\;\frac{\sin(2\pi\tilde\omega/\Omega_f)L} {2\pi\tilde\omega
/\Omega_f}|\omega_p/2+\tilde\omega\rangle_s
|\omega_p/2-\tilde\omega\rangle_i
\;\labell{tbstate}
\end{eqnarray}
where \begin{eqnarray}
\Omega_f=\frac {4\pi}{L|\gamma_s-\gamma_i|}
\;\labell{bandw},  \end{eqnarray} and a normalization constant has
been omitted. In this limit, the fluorescence spectra of the two
photons are identical and have a bandwidth $\Omega_f$. In the case of
type I phase-matching (for which $\gamma_s=\gamma_i$) the derivation
of Eq.~(\ref{pmstato}) would require one to include second-order terms
in the expansion (\ref{deltaktaylor}), because the first-order terms
cancel {\cite{walms}}. In the following, for sake of simplicity, we
limit our analysis to type II crystals for which the linear
approximation is sufficient. The biphoton twin-beam state $|{\rm
TB}\rangle$ is a maximally frequency-entangled state in the sense that
a frequency measurement on one photon exactly determines the outcome
of a frequency measurement on the other photon. In particular, the
frequencies of the signal and idler are anti-correlated: their sum is
fixed and equal to the pump frequency.

Two experimental problems, connected with a finite pump bandwidth
$\Omega_p$ and a finite crystal length $L$, arise when the
conventional phase-matching condition is used to generate
entanglement. If the pump spectrum is not perfectly monochromatic
($\Omega_p>0$), a range of signal-idler sum frequencies exist such
that the maximal entanglement property is degraded and the spectra of
the two down-converted photons are no longer identical
{\cite{wal}}. The drawback of non-identical signal and idler spectra
turns out to be critical for many experiments that employ a pulsed
pump, because the difference in the spectra introduces
distinguishability that is detrimental to quantum interference
{\cite{keller,wal,ata,theory}}. On the other hand, it is clear from
Eq.~(\ref{bandw}) that the use of long crystals reduces the available
bandwidth over which frequency entanglement occurs.  In the limit of a
long crystal no entanglement survives because signal and idler are
both monochromatic photons of frequency $\omega_p/2$. However, from
the experimental point of view, a long crystal is preferable because
it increases down-conversion efficiency. Various experimental
techniques aimed at preserving or recovering entanglement in the
pulsed-pump regime have been studied {\cite{pulse}}. In this paper, we
approach these problems by extending the conventional phase-matching
condition (\ref{ordine0}) in such a way that the signal and idler
spectra remain indistinguishable for any pump bandwidth.

\section{Extended phase-matching conditions}\labell{s:ext}
Under the conventional phase-matching condition for
frequency-degenerate SPDC, and in the linear dispersion regime of
Eq.~(\ref{deltaktaylor}), we have shown that the phase-matching
function $\Phi_L$ is characterized by the two parameters $\gamma_s$
and $\gamma_i$. Better physical insight may be gained by treating
$\gamma_s$ and $\gamma_i$ as Cartesian coordinates and then converting
to the polar-coordinates representation $\gamma_s=\gamma\cos\theta$
and $\gamma_i=\gamma\sin\theta$. The parameter $\theta$ controls the
structure of the biphoton spectrum by governing the orientation of the
symmetry axis $\tilde\omega_i=-\tilde\omega_s\tan\theta$ of the
phase-matching function $\Phi_L(\omega_s,\omega_i)$. The parameter
$\gamma$, on the other hand, controls the width of $\Phi_L$ and hence
determines the bandwidth of the biphoton state. As will be shown, it
is the parameter $\theta$ that effectively controls the ``quality'' of
the frequency entanglement of the down-converted photons, thus
determining their frequency correlations.

To enforce the indistinguishability of the down-converted signal and
idler spectra, it is necessary to symmetrize the biphoton spectral
amplitude $A(\omega_s,\omega_i)$ with respect to $\omega_s$ and
$\omega_i$ {\cite{wal}}.  Given that $\alpha(\omega_s,\omega_i)$ of
Eq.~(\ref{alpha}) can be approximated by a function of $\omega_s +
\omega_i$, Eq.~(\ref{defa}) suggests that we need only to symmetrize
the phase-matching function $\Phi_L(\omega_s,\omega_i)$ to guarantee
that $A(\omega_s,\omega_i)$ is symmetric in $\omega_s$ and
$\omega_i$. The simplest and best known way to achieve such symmetry
is to require type I phase-matched crystals for which
$k_s(\omega)=k_i(\omega)$. Interestingly, a whole new class of states
is obtained by adopting a different strategy: the symmetrization of
$\Phi_L$ can also be achieved by requiring $\gamma_s=-\gamma_i$
({i.e.}  $\theta=-\pi/4$). This regime can be enforced if the crystal,
in addition to satisfying the conventional phase-matching condition
(\ref{ordine0}), also satisfies the group velocity condition
{\cite{keller,walms,noi}}
\begin{eqnarray}
k'_p(\omega_p) =
\frac{k'_s(\omega_p/2)+k'_i(\omega_p/2)}2\;.
\labell{epm}
\end{eqnarray}
Equations (\ref{ordine0}) and (\ref{epm}) force the system to be phase
matched over the entire fluorescence spectrum, not just at
degeneracy. Together they constitute the extended phase-matching
conditions that will be used in the rest of this paper {\footnote{When
type I phase-matched crystals are employed, condition (\ref{epm})
implies $\gamma_s=\gamma_i=0$. In this case, as already noticed,
Eq.~(\ref{deltaktaylor}) is no longer valid and we need to include
higher-order contributions {\cite{walms}}. The limits of validity of
the first-order Taylor expansion and the generalization of
Eq.~(\ref{epm}) to higher orders will be discussed at the end of
Sect.~\ref{s:ext}.  }.  Under these constraints, the state
$|\Psi\rangle$ now becomes
\begin{widetext}
\begin{eqnarray} &&|\Psi_{epm}\rangle=
\int\! \frac{\displaystyle d\tilde\omega_s}{\displaystyle 2\pi}
\int\! \frac{\displaystyle d\tilde\omega_i}{\displaystyle 2\pi}\;
\alpha(\tilde\omega_s+\tilde\omega_i+\omega_p)\;\labell{epmstato}
\frac{\sin\left[\pi(\tilde\omega_s-\tilde\omega_i)/\Omega_f\right]L}
{\pi(\tilde\omega_s-\tilde\omega_i)/\Omega_f}\;
|\omega_p/2+\tilde\omega_s\rangle_s\;|\omega_p/2+\tilde\omega_i\rangle_i
\;.
\end{eqnarray}
\end{widetext}
The biphoton spectral amplitude of this state is of the form
\begin{eqnarray}
A(\omega_s,\omega_i)=S(\omega_s+\omega_i)\;D(\omega_s-\omega_i)
\;\labell{somdiff},
\end{eqnarray}
where $S\equiv\alpha$ is a function of the frequency sum
$\omega_s+\omega_i$, and $D\equiv\Phi_L$ is a function of the
frequency difference $\omega_s-\omega_i$. The symmetry properties of
the extended-phase-matched $A(\omega_s,\omega_i)$ are such that,
independent of the pump bandwidth $\Omega_p$, the symmetry axis of the
pump spectral function $\alpha(\omega_s+\omega_i)$ (with
$\theta=\pi/4$) is orthogonal to the symmetry axis of the phase-matching function $\Phi_L(\omega_s-\omega_i)$ (with $\theta=-\pi/4$).
In contrast, the symmetry axis of $\Phi_L(\omega_s,\omega_i)$ for a
conventionally phase-matched state $|\Psi_{pm}\rangle$ is not
specified and can take on any value of $\theta$ (see
Fig.~\ref{f:stellina}).

\begin{figure}[hbt]
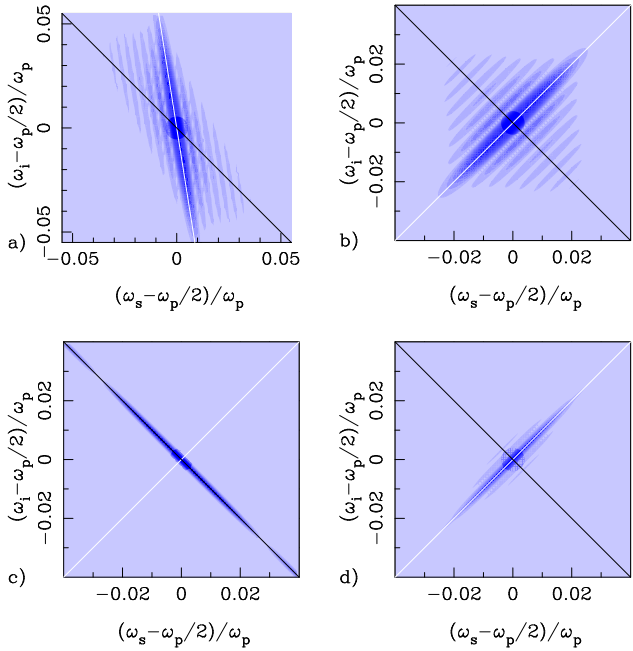

\begin{center}
\epsfxsize=.45\hsize\leavevmode\epsffile{figure1a.eps}
\hspace{.3cm}
\epsfxsize=.45\hsize\leavevmode\epsffile{figure1b.eps}
\end{center}
\begin{center}
\epsfxsize=.45\hsize\leavevmode\epsffile{figure1c.eps}
\hspace{.3cm}
\epsfxsize=.45\hsize\leavevmode\epsffile{figure1d.eps}
\end{center}
\caption{Plots of the biphoton spectral amplitude
$|A(\omega_s,\omega_i)|$ for different values of the crystal length
$L$ and pump bandwidth $\Omega_p$. In each plot the white line is the
symmetry axis of $\Phi_L(\omega_s,\omega_i)$ and the black line is the
symmetry axis of $\alpha(\omega_s+\omega_i)$, which is chosen
according to (\ref{gaus}).  {\bf a)}~Conventional phase-matching
condition (\ref{ordine0}): the two symmetry axes are in general
neither orthogonal nor coincident. (the parameters for this plot are
$\theta=\pi/20$; $L=1\,$cm; $\Omega_p=4\times 10^{13}$\,s$^{-1}$.) 
{\bf b)}~Extended phase-matching conditions (\ref{ordine0}) and
(\ref{epm}): the two symmetry axes are orthogonal. ($\theta=-\pi/4$;
$L=1\,$cm; $\Omega_p=4\times 10^{13}$\,s$^{-1}$.)  {\bf c)}~$|{\rm
TB}\rangle$-like state (\ref{tbstate}): signal and idler photon
frequencies are strongly anti-correlated. The $|{\rm TB}\rangle$ state
is reached by using either the conventional phase-matching condition
or the extended ones in the limit of small bandwidth
$\Omega_p$. ($L=0.1\,$cm; $\Omega_p=1.6\times 10^{12}$\,s$^{-1}$.) 
{\bf d)}~$|{\rm DB}\rangle$-like state (\ref{dbstate}): signal and
idler frequencies are strongly positive-correlated. The $|{\rm
DB}\rangle$ state can {\it only} be obtained under the extended
phase-matching conditions in the limit of very long
crystals. ($L=5\,$cm; $\Omega_p=4\times 10^{13}$\,s$^{-1}$.) For all
the plots $\omega_p=2\times 10^{15}\,$s$^{-1}$ and $\gamma=8\times
10^{-5}\,$ps/$\mu$m. }
\labell{f:stellina}\end{figure}

In the limit of a monochromatic pump, $\Omega_p\to 0$, we again obtain
the twin-beam state $|{\rm TB}\rangle$ from $|\Psi_{epm}\rangle$,
suggesting that the additional phase-matching constraint
Eq.~(\ref{epm}) does not play a role in cw-pumped SPDC\@. On the other
hand, even for pulsed pump, we may still obtain a maximally
frequency-entangled state from $|\Psi_{epm}\rangle$ in the limit of
infinite crystal length $L\to\infty$ {\cite{walms,noi}}. In this case
one obtains the ``difference-beam'' state in which the signal and
idler frequencies are equal, {i.e.}\begin{eqnarray} |{\rm
DB}\rangle\equiv\int
\frac{\displaystyle d\tilde\omega}{\displaystyle
2\pi}\;\alpha(2\tilde\omega+\omega_p)|\omega_p/2+\tilde\omega\rangle_s
|\omega_p/2+\tilde\omega\rangle_i
\;\labell{dbstate},
\end{eqnarray}
where the spectral function of the state is now determined entirely by
the pump spectral characteristics. As shown pictorially in
Fig.~\ref{f:fotoni}, the properties of this $|{\rm DB}\rangle$ state
are complementary, via Fourier duality, to those of the $|{\rm
TB}\rangle$ state. A detailed comparison between the states $|{\rm
TB}\rangle$ and $|{\rm DB}\rangle$ can be found in {\cite{noi}}. In
this paper we will focus on the properties of the class of states
$|\Psi_{epm}\rangle$ of Eq.~(\ref{epmstato}) to which both $|{\rm
TB}\rangle$ and $|{\rm DB}\rangle$ belong.

\begin{figure}[hbt]
\begin{center}\epsfxsize=.75
\hsize\leavevmode\epsffile{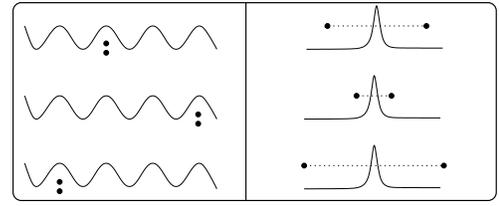}
\end{center}
\caption{Pictorial representation of the time domain description of
parametric down-conversion. In the crystal, the signal and idler
photons are created simultaneously within a coherence time of the
pump. {\bf Left box:} In the $|{\rm TB}\rangle$ state generation, a cw
pump (represented by the sine wave in the figure) is used and its
(ideally infinite) coherence time is much longer than the average time
that the generated photons need to traverse the crystal ($\sim
L/c$). The two photons in the $|{\rm TB}\rangle$ state are {\it time
correlated} and, by Fourier duality, {\it frequency
anti-correlated}. {\bf Right box:} In the $|{\rm DB}\rangle$ state
generation, a narrow pump pulse and a long (ideally infinite)
nonlinear crystal are required. In this case, the different dispersion
that the two photons see while crossing the long nonlinear crystal
after their generation tends to separate them. At the output of the
crystal their separation depends on the position at which they were
created. The extended phase-matching conditions, however, guarantee
that the mean ``position'' of the two photons coincides with that of
the pump pulse. Hence, the two photons in the $|{\rm DB}\rangle$ state
are {\it time anti-correlated} and {\it frequency correlated}.  }
\labell{f:fotoni}\end{figure}

The states $|\Psi_{epm}\rangle$ comprise a family of
frequency-entangled states each of which can be uniquely identified by
the value of the two bandwidth parameters $\Omega_f$ of
Eq.~(\ref{bandw}) and $\Omega_p$. The parameter $\Omega_f$ derives
from the phase-matching function $\Phi_L$ and is determined by the
crystal properties alone. It sets the bandwidth of the
frequency-difference part of the biphoton state. On the other hand,
the parameter $\Omega_p$, comes solely from the pump spectral function
$\alpha$ and determines the bandwidth of the frequency-sum part of the
biphoton state.  The ideal cases of $|{\rm TB}\rangle$ and $|{\rm
DB}\rangle$ are the maximally entangled extrema of this family of
states. For finite $L$ and nonzero $\Omega_p$, the state
$|\Psi_{epm}\rangle$ retains frequency entanglement even though it is
no longer maximally entangled. Hence, a measurement of the frequency
of one photon partially determines the frequency of the
other. Nevertheless, the spectra of the two photons are the same and
the state maintains optimal visibility in quantum interference
experiments for all values of the pump bandwidth $\Omega_p$. This
aspect will be thoroughly analyzed in Sect.~\ref{s:car}, where some
interferometric measurements to characterize these states are
described.

The extended phase-matching conditions (\ref{ordine0}) and (\ref{epm})
have been obtained from the first-order Taylor-series expansion given
in Eq.~(\ref{deltaktaylor}).  This approximation for the function
$\Phi_L$ of Eq.~(\ref{phi}) is valid if the second-order correction to
$\Delta k(\omega_s,\omega_i)$ introduces a contribution $\delta_k$
such that $\delta_kL/2$ is much smaller than $\pi/2$.  In vector
notation, this second-order term is given by
$\frac12(\vec\omega-\vec\omega_0)\cdot
H(\vec\omega_0)\cdot(\vec\omega-\vec\omega_0)$ where
$\vec\omega=(\omega_s,\omega_i)$,
$\vec\omega_0=(\omega_p/2,\omega_p/2)$, and ${H}({\vec\omega_0})$ is
the Hessian matrix of the function $\Delta k(\vec\omega)$ evaluated at
$\vec\omega_0$. The second-order contribution can be shown to be
smaller than $|\mu|\Omega_p^2/2$, where $\mu$ is the maximum-magnitude
eigenvalue of ${H}(\vec{\omega_0})$. Hence, to ensure the validity of
the first-order expansion in (\ref{deltaktaylor}) over the entire pump
spectrum $\Omega_p$, we must require that $L\ll
8\pi/(|\mu|\Omega_p^2)$. An example of a nonlinear crystal that
satisfies this requirements for type II extended phase-matching has
been given in {\cite{noi}}.

When the second-order contribution cannot be neglected, we can still
obtain a biphoton spectral function of the form (\ref{somdiff}). This
can be achieved by augmenting (\ref{ordine0}) and (\ref{epm}) with the
requirement that
\begin{eqnarray}
k''_s(\omega_p/2)=k''_i(\omega_p/2)=2k''_p(\omega_p)
\;\labell{secord}.
\end{eqnarray}
If contributions of order greater than or equal to three are
non-negligible, then there is no way to impose the form of
Eq.~(\ref{somdiff}) to the spectral function
$A(\omega_s,\omega_i)$. It is still possible, however, to force the
maximum of the phase-matching function $\Phi_L(\omega_s,\omega_i)$ to
lie in the region $\omega_s=\omega_i$, as in the case of the $|{\rm
DB}\rangle$ state. This is obtained by requiring that the $n$th order
terms obey
\begin{eqnarray}
{\frac{\partial^nk_{p}}{\partial\omega^n}(\omega_p)=\frac 1{2^n}
\left[
\frac{\partial^nk_{s}}{\partial\omega^n}(\omega_p/2)+
\frac{\partial^nk_{i}}{\partial\omega^n}(\omega_p/2)\right]\;.} 
\;\labell{genepm}
\end{eqnarray}
In the rest of the paper we will focus on cases in which the
first-order approximation holds.

\section{Polarization entanglement}\labell{s:pol}
Because the two photons of $|\Psi_{epm}\rangle$ are indistinguishable
in frequency, it would be very useful to entangle them in
polarization.  A number of configurations can be utilized to achieve
this goal.  For example one can adopt a method analogous to the one
presented in {\cite{jeff}}, in which the outputs from two
coherently-pumped optical parametric amplifiers (OPAs) are entangled
with a polarizing beam splitter. Here, we discuss a different scheme,
based on the one presented in {\cite{singleopa}}, that
employs a single crystal. It generates the desired
polarization-entangled state only $50\%$ of the time, but a simple
post-selection measurement allows one to discard the cases in which
the polarization-entangled biphoton state is not present.

In a type II phase-matched crystal the down-converted photons have
orthogonal polarizations, say $\updownarrow$ for the signal and
$\leftrightarrow$ for the idler. If these photons are fed into one
port of a 50-50 beam splitter, then each of the two photons (which are
distinguishable in polarization) has an equal chance of being
transmitted or reflected. The state after the beam splitter is given
by \begin{eqnarray} &&|\Psi_{out}\rangle=
\frac 12\int \frac{d\omega_s}{2\pi} \int \frac 
{d\omega_i}{2\pi}\; A(\omega_s,\omega_i)\;
[c^\dag_\updownarrow(\omega_s)
b^\dag_\leftrightarrow(\omega_i)+
\nonumber\\&&
b^\dag_\updownarrow(\omega_s)
c^\dag_\leftrightarrow(\omega_i)+b^\dag_\updownarrow(\omega_s)
b^\dag_\leftrightarrow(\omega_i)+c^\dag_\updownarrow(\omega_s)
c^\dag_\leftrightarrow(\omega_i)]|0\rangle\labell{psiout},
\end{eqnarray} 
where $b$ and $c$ are the output modes of the beam splitter. At the
two output ports of the beam splitter, detectors with measurement
intervals longer than the largest signal--idler time separation are
used to measure the desired coincidences even though the two photons
will not in general arrive exactly at the same time {\cite{noi}}. This
post-selection scheme discards the cases in which both photons of a
pair exit from the same beam-splitter output port, {i.e.}, the last
two terms of Eq.~(\ref{psiout}). The post-selected, filtered output is
the Bell state
\begin{eqnarray} 
|\Psi^{(+)}\rangle=\frac
1{\sqrt{2}}\Big[|\!\updownarrow\rangle|\!\leftrightarrow\rangle+
|\!\leftrightarrow\rangle|\!\updownarrow\rangle\Big]\;.
\labell{bellst}
\end{eqnarray} 
The other three Bell states can be similarly obtained by appropriate
addition of a polarization rotator and/or a $\pi$-rad phase shifter
acting on one of the polarizations in one of the beam splitter output
arms.

\section{Experimental entanglement characterization}\labell{s:car}
In this section we show how the frequency entanglement of the family
of states $|\Psi_{epm}\rangle$ can be characterized
experimentally. The setups we describe are sketched in
Fig.~\ref{f:experim} and are based on the Hong-Ou-Mandel (HOM)
interferometer {\cite{manou}} and on the Mach-Zehnder (MZ)
interferometer {\cite{mz}}. The photodetectors 1 and 2 at the output
of the interferometers measure the photo-coincidence rate $P(\tau)$
over a long-time ({i.e.}, longer than the time duration of the
biphoton) detection window. The rate $P$ is monitored for different
values of the time delay $\tau$ between the two arms of the
interferometers that can be varied by moving the 50-50 output beam
splitter (BS). From the Mandel formula for photodetection
{\cite{mandel}}, we find 
\begin{eqnarray}
P&\propto&\int_Tdt_1\int_Tdt_2\;\langle\Psi_{in}|E_1^{(-)}(t_1)
E_2^{(-)}(t_2)\nonumber\\&\times&
E_2^{(+)}(t_2)E_1^{(+)}(t_1)|\Psi_{in}\rangle
\;\labell{mand},
\end{eqnarray}
where $|\Psi_{in}\rangle$ is the state of the field at the
interferometer input, $T$ is the measurement interval of the detectors
and $E^{(\pm)}_j$ refers to the negative and positive components of
the electric field at the $j$th detector.  When $|\Psi_{in}\rangle$ is
a biphoton state such as Eq.~(\ref{stato}), the expression
(\ref{mand}) in the limit of long detection window $T\to\infty$
becomes \begin{eqnarray} P\propto \int
\!\frac{\displaystyle d\omega_1}{\displaystyle 2\pi}
\int \!\frac{\displaystyle d\omega_2}{\displaystyle 2\pi}\,|\langle
0|a_1(\omega_1)
a_2(\omega_2)|\Psi_{in}\rangle|^2,
\;\labell{rate}
\end{eqnarray}
where $a_1$ and $a_2$ are the photon annihilation operators at the two
detectors. In terms of the annihilation operators of the signal and
idler, these are given by
\begin{eqnarray}
a_1(\omega)&=&\left[
a_s(\omega)\;e^{i\omega\tau}+a_i(\omega)\right]/\sqrt{2}\;\labell{homtr}
\\
\nonumber
a_2(\omega)&=&\left
[a_s(\omega)-a_i(\omega)\;e^{-i\omega\tau}\right]/{\sqrt{2}}
\end{eqnarray}
for the HOM interferometer setup and by
\begin{eqnarray}
a_1(\omega)&=&\left[a_s(\omega)
(e^{i\omega\tau}+1)+a_i(\omega)(e^{i\omega\tau}-1)
\right]/2\;\labell{mztr}
\\
\nonumber a_2(\omega)&=&\left[a_s(\omega)
(1-e^{-i\omega\tau})+a_i(\omega)(1+e^{-i\omega\tau})\right]/2
\end{eqnarray}
for the MZ interferometer setup.

\begin{figure}[hbt]
\begin{center}\epsfxsize=.75
\hsize\leavevmode\epsffile{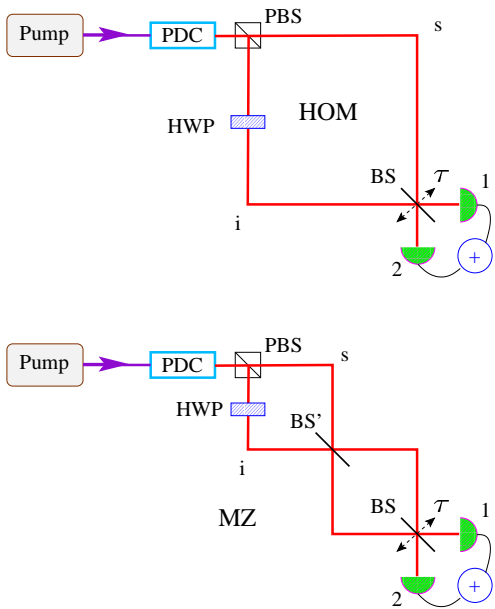}
\end{center}
\caption{Schematics of proposed experiments. The upper setup
implements a Hong-Ou-Mandel interferometer that (under the extended
phase-matching conditions) is sensitive to frequency anti-correlation
of the down-converted photons. The lower setup is a Mach-Zehnder
interferometer that is sensitive to positive frequency correlation. In
both experiments the coincidences at detectors 1 and 2 are measured
for different values of the relative delay $\tau$ between the two
interferometer arms. The pump source produces a coherent pulse of mean
frequency $\omega_p$ and bandwidth $\Omega_p$ that can be varied to
create different types of entanglement. The configurations shown refer
to type II phase-matched parametric down-converter (PDC) crystals for
which a polarizing beam splitter (PBS) separates the signal and idler
photons and the half wave plate (HWP) guarantees that the
polarizations in the two output beams are the same.}
\labell{f:experim}\end{figure}

If the biphoton spectral amplitude $A(\omega_s,\omega_i)$ is symmetric
in its arguments, as in the case of $|\Psi_{epm}\rangle$,
Eq.~(\ref{rate}) becomes
\begin{eqnarray}
P_{\pm}(\tau)&\propto&\int \! \frac{\displaystyle d\omega_1}{\displaystyle
2\pi}\int \!
\frac{\displaystyle d\omega_2}{\displaystyle
2\pi}\,|A(\omega_1,\omega_2)|^2\nonumber\\&\times &
\Big(1\pm
\cos[(\omega_1\pm\omega_2)\tau]\Big),
\labell{biancaneve}
\end{eqnarray}
where the minus signs apply to the HOM interferometer, and the plus
signs apply to the MZ interferometer. Using Eq.~(\ref{somdiff}), one
can show that the two interferometers are sensitive to different parts
of the biphoton spectrum:
\begin{eqnarray}
P_-(\tau)
&\propto&\int
\frac{d\omega}{2\pi}\;|D(\omega)|^2\left[1-\cos(\omega\tau)\right]\;,
\labell{hom}
\\P_+(\tau)&\propto&
\int \frac{d\omega}{2\pi}\;|S(\omega)|^2\left[1+\cos(\omega\tau)
\right]\;,
\labell{mz1}
\end{eqnarray}
{i.e.}, the HOM measures the difference-frequency part of the biphoton
spectrum of the state $|\Psi_{epm}\rangle$, and the MZ measures the
sum-frequency part.

\begin{figure}[hbt]
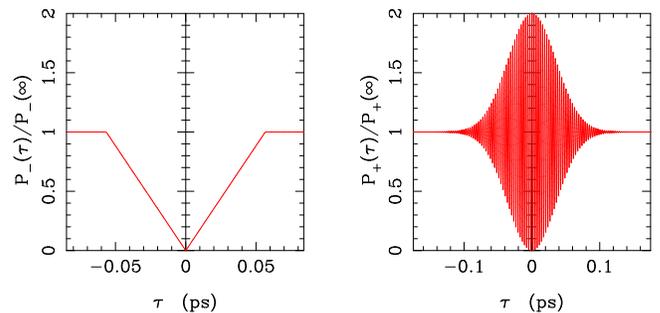

\begin{center}\epsfxsize=.45
\hsize\leavevmode\epsffile{figure4a.eps}
\hspace{.5 cm}
\epsfxsize=.45
\hsize\leavevmode\epsffile{figure4b.eps}
\end{center}
\caption{Example of quantum signatures of the experimental
results for a $|\Psi_{epm}\rangle$ input state. {\bf Left:} Mandel dip
in the coincidence rate $P_-(\tau)$ at the output of a HOM
interferometer. The characteristic triangular shape of width
$\propto1/\Omega_f$ comes from the $\sin x/x$ form of the biphoton
spectrum in (\ref{epmstato}). {\bf Right:} Fringes and fringe envelope
in the coincidence rate $P_+(\tau)$ at the output of the MZ
interferometer. The Gaussian envelope of width $\propto1/\Omega_p$
derives from the Gaussian spectral profile (\ref{gaus}) of the pump
beam. The parameters in the plots are $\gamma L=8\times 10^{-2}\ $ps,
$\omega_p=2\times 10^{15}$\,s$^{-1}$ and $\Omega_p=4\times
10^{13}$\,s$^{-1}$.}
\labell{f:dip}\end{figure}

Consider first the HOM interferometer. Equation (\ref{hom}) yields the
usual triangular Mandel dip for type II phase-matching (shown in
Fig.~\ref{f:dip}) centered at $\tau=0$ with a (base-to-base) width of
$4\pi/\Omega_f$, which is determined by the crystal length $L$. Note
that the pump bandwidth $\Omega_p$ does not play any role in
determining the function $P_-(\tau)$, which is a unique characteristic
of the state $|\Psi_{epm}\rangle$ because of the extended
phase-matching conditions. It is well known {\cite{keller,wal,ata}}
that, under the conventional phase-matching condition (\ref{ordine0})
only, the visibility of the Mandel dip decreases when the pump
bandwidth is increased. The Mandel dip can be interpreted as the
result of destructive quantum interference between the different paths
that lead to coincidences at the detectors {\cite{pittman}}. If we are
limited to the conventional phase-matching condition in type II
phase-matched crystals, the two down-converted photons are
distinguishable (for $\Omega_p>0$) because their spectra are different
{\cite{wal}}. In contrast, the extended phase-matching conditions
ensure that the two spectra are equal (independent of $\Omega_p$),
thus maintaining indistinguishability of the two photons for all
values of $\Omega_p$. In this case, the visibility is not lost when
the pump bandwidth $\Omega_p$ is increased {\cite{walms}}. This is
evident from Fig.~\ref{f:visib}, where the visibilities of experiments
employing different phase-matching conditions are compared. A detailed
analysis of the visibility is presented in App.~\ref{s:app}.

\begin{figure}[hbt]
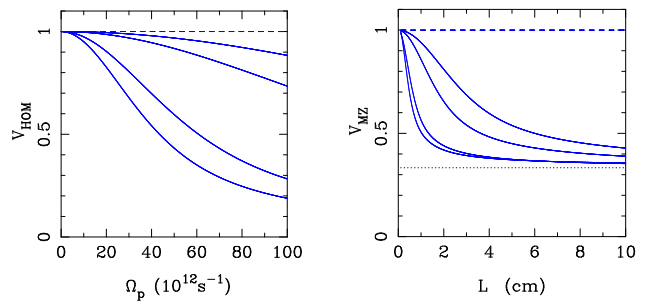

\begin{center}\epsfxsize=.45
\hsize\leavevmode\epsffile{figure5a.eps}
\hspace{.5 cm}\epsfxsize=.45
\hsize\leavevmode\epsffile{figure5b.eps}
\end{center}
\caption{{\bf Left:} Plot of the  Mandel-dip visibility
$V_{HOM}$ of Eq.~(\ref{vhom}), which measures the dip depth in the HOM
interferometer as a function of the pump bandwidth $\Omega_p$. Here
$\gamma=8\times 10^{-5}$\,ps/$\mu$m; $L=0.1$\,cm. {\bf Right:} Plot of
the MZ-peak visibility $V_{MZ}$ of Eq.~(\ref{caleffe}), which measures
the amplitude of the modulation of the fringes of Fig.~\ref{f:dip} as
a function of the crystal length $L$. As discussed in
appendix~\ref{s:app}, the MZ visibility is always greater than $33\%$
(dotted line). Here $\gamma=8\times 10^{-5}$\,ps/$\mu$m;
$\Omega_p=4\times 10^{13}$\,s$^{-1}$.  For both plots different curves
refer to different phase-matching conditions: extended phase-matching
condition $\theta=-\pi/4$ (upper dashed line); conventional
phase-matching condition (lower solid lines for, from top to bottom,
$\theta=-\pi/5$, $\theta=-\pi/6$, $\theta=0$, $\theta=\pi/5$). In both
the HOM and MZ cases, under the conventional phase-matching condition
the visibility decreases for increasing pump bandwidth
{\cite{keller,wal,ata}} or crystal length. In contrast, there is no
loss of visibility when the extended phase-matching conditions
($\theta=-\pi/4$) are employed. }
\labell{f:visib}\end{figure}

Consider now the MZ interferometer. Equation (\ref{mz1}) gives a peak
centered at $\tau=0$ with a width that is proportional to $1/\Omega_p$
and is determined by the pump spectrum. This peak is modulated by
fringes of frequency $\omega_p$ and is totally independent of the
crystal properties. An example is given in Fig.~\ref{f:dip}. For
Gaussian pump spectrum
\begin{eqnarray}
|\alpha(\omega)|^2\propto
\exp[-(\omega-\omega_p)^2/\Omega_p^2]
\;\labell{gaus},
\end{eqnarray}
the coincidence function becomes \begin{eqnarray} P_+(\tau)\propto
1+\exp(-\Omega_p^2\tau^2/4)\cos(\omega_p\tau)
\;\labell{pipiumz}.
\end{eqnarray}
Similar fringes in coincidence counts have been experimentally
observed under the conventional phase-matching condition in
{\cite{mz}}. Analogous to the case of the HOM interferometer, if one
drops the extended phase-matching conditions in favor of the
conventional phase-matching condition (\ref{ordine0}), the visibility
of the MZ peak decreases when $L$ is increased. This effect is derived
in App.~\ref{s:app} and illustrated in Fig.~\ref{f:visib}. Similar to
the HOM dip, it is possible to interpret the loss of visibility in
terms of loss of indistinguishability of the two down-converted
photons.

As shown in Sect.~\ref{s:ext}, each element of the family of states
can be uniquely determined once the phase-matching bandwidth
$\Omega_f$ of Eq.~(\ref{bandw}) and the pump bandwidth $\Omega_p$ are
specified. It follows from the above analysis that the HOM
interferometer cannot distinguish among $|\Psi_{epm}\rangle$ states
with a fixed $\Omega_f$ but different $\Omega_p$. It can only be used
to discriminate among states with different $\Omega_f$. On the other
hand, the MZ interferometer cannot distinguish among states with a
fixed $\Omega_p$ but different $\Omega_f$. It can only discriminate
among states with different pump bandwidths $\Omega_p$.  In order to
obtain a complete characterization of the family of states
$|\Psi_{epm}\rangle$, one needs to perform both HOM and MZ
interferometric experiments. However, if one limits the analysis to
the extremal cases of the states $|{\rm TB}\rangle$ and $|{\rm
DB}\rangle$, either one of the two experiments is sufficient to
distinguish between the two states because they are generated with
different pump bandwidths {\it and} different crystal lengths.

Before concluding, some final remarks are in order. The unique
characteristics of the $|\Psi_{epm}\rangle$ states originate from the
symmetry properties imposed by the extended phase-matching
conditions. These conditions allow the quantum correlations of the
state to be more resilient. However, even for the conventionally
phase-matched state $|\Psi_{pm}\rangle$ analogous properties could be
obtained, if one were able to devise appropriate interferometric
setups that somehow compensate for the difference between the signal
and idler spectra ({e.g.}, by stretching the bandwidth of one of
them).  In this respect we can say that the fundamental difference
between the $|\Psi_{epm}\rangle$ and $|\Psi_{pm}\rangle$ states is in
the fact that the former requires simpler and more easily realizable
setups for demonstrating nonclassical features of the quantum states.

\section{Conclusions}\labell{s:concl}
We have discussed spontaneous parametric down-conversion in the
regimes of continuous-wave pumping and pulsed pumping.  By adding to
the conventional crystal phase-matching condition (\ref{ordine0}) the
constraint (\ref{epm}), it is possible to enforce certain symmetry
properties on the biphoton spectral function
$A(\omega_s,\omega_i)$. This allows one to create a down-converted
pair of photons with identical spectra that, under accessible
experimental conditions, constitutes a maximally-entangled biphoton
whose component photons have coincident frequencies.  We have proposed
and analyzed two experimental arrangements, based on the HOM and MZ
interferometers, that can be used to characterize the states that exit
an extended-phase-matched crystal. In particular, we have shown that
extended-phase-matched states retain maximal visibility on both the
interferometers for all the pump spectra. As in the case of the
conventional phase-matching, it is possible to create polarization
entanglement by exploiting the frequency entanglement: we have
described a post-selective entangling procedure.

\appendix
\section{Visibility of HOM and MZ interferometers}\labell{s:app}
In this appendix we derive the expressions for the visibility
functions $V_{HOM}$ and $V_{MZ}$ of the two experiments that were
discussed in Sect.~\ref{s:car}.

We have already analyzed extended-phase-matched crystals. In the case
of conventionally phase-matched crystals, where Eq.~(\ref{epm}) does
not hold, the expressions for the photo-coincidence rates
$P_\pm(\tau)$ of Eqs.~(\ref{hom}) and (\ref{mz1}) are no longer
valid. However, starting from Eq.~(\ref{rate}), one can obtain the
rate for the Hong-Ou-Mandel interferometer as
\begin{eqnarray}
&&
P_-(\tau)\propto\int\frac{d\tilde\omega_1}{2\pi}
\int\frac{d\tilde\omega_2}{2\pi}
|\alpha(\tilde\omega_1+\tilde\omega_2+\omega_p)|^2
\labell{pmen}\\\nonumber&&
\left|
\phi_L(\gamma_s\tilde\omega_2+\gamma_i\tilde\omega_1)-
\phi_L(\gamma_s\tilde\omega_1+\gamma_i\tilde\omega_2)\;
e^{i(\tilde\omega_1-\tilde\omega_2)\tau}\right|^2
\;,
\end{eqnarray}
where $\phi_L(x)=2\sin(xL/2)/x$. For a Gaussian pump spectrum
$|\alpha(\omega)|^2$ of the form (\ref{gaus}), Eq.~(\ref{pmen})
reduces to \begin{eqnarray}
&&P_-(\tau)\propto\left\{\matrix{1-\frac{\sqrt{\pi}}2\xi\;\mbox{Erf}
\left(\frac{\displaystyle 1-|\tau|/\tau_\theta}{\displaystyle\xi}
\right)
&\mbox{ for }|\tau|<\tau_\theta\cr\cr 1&\mbox{ for }
|\tau|>\tau_\theta&\!\!\!,}\right.
\nonumber\\&&\;\labell{brontolo}
\end{eqnarray}
where $\mbox{Erf}(x)=\frac 2{\sqrt{\pi}}\int_0^xdy\;e^{-y^2}$ is the
error function and \begin{eqnarray}
\xi&\equiv&4/\Big({\Omega_p\gamma L|\cos\theta+\sin\theta|}\Big)
\nonumber\\
\tau_\theta&\equiv&{\gamma L|\cos\theta-\sin\theta|}/2
\;\labell{cucciolo}.
\end{eqnarray}
For $\xi\to\infty$, $P_-(\tau)$ reduces to the familiar
triangular-shaped Mandel dip plotted in Fig.~\ref{f:dip}. This limit
can be achieved for vanishing pump bandwidth $\Omega_p\to 0$,
i.e. when a $|{\rm TB}\rangle$ state is fed into the
interferometer. Another way to obtain the same limit is to employ the
extended phase-matching conditions ($\theta=-\pi/4$). This implies
that an extended-phase-matched crystal will always show perfect
visibility for any value of the pump bandwidth $\Omega_p$. The
visibility is defined as the depth of the dip of Eq.~(\ref{brontolo}),
i.e. \begin{eqnarray}
V_{HOM}\equiv\frac{P_-(\infty)-P_-(0)}{P_-(\infty)+P_-(0)}=
\frac{\frac{\sqrt{\pi}}2\xi\;\mbox{Erf}\left(1/\xi\right)}
{2-\frac{\sqrt{\pi}}2\xi\;\mbox{Erf}\left(1/\xi\right)}
\;\labell{vhom}.
\end{eqnarray}
It is plotted as a function of $\Omega_p$ for different values of
$\theta$ in Fig.~\ref{f:visib}.

The case of the Mach-Zehnder interferometer is analogous, but
now Eq.~(\ref{rate}) becomes
\begin{widetext}
\begin{eqnarray} 
&&P_+(\tau)\propto\int\frac{d\tilde\omega_1}{2\pi}
\int\frac{d\tilde\omega_2}{2\pi}
|\alpha(\tilde\omega_1+\tilde\omega_2+\omega_p)|^2
\labell{pipi}
\\\nonumber&&\times
\Big|\phi_L(\gamma_s\tilde\omega_2+\gamma_i\tilde\omega_1)
\sin\left[(\tilde\omega_1+\frac{\omega_p}2)\frac\tau 2\right]
\sin\left[(\tilde\omega_2+\frac{\omega_p}2)\frac\tau 2\right]
-\phi_L(\gamma_s\tilde\omega_1+\gamma_i\tilde\omega_2)
\cos\left[(\tilde\omega_1+\frac{\omega_p}2)\frac\tau 2\right]
\cos\left[(\tilde\omega_2+\frac{\omega_p}2)\frac\tau 2\right]\Big|^2
\;.
\end{eqnarray}
Considering again a Gaussian pump spectrum, Eq.~(\ref{pipi}) reduces
to
\begin{eqnarray} P_+(\tau)\propto
1+\cos(\omega_p\tau){\cal F}_1(\tau)+{\cal F}_2(\tau)
\;\labell{pipimeglio},
\end{eqnarray}
where 
\begin{eqnarray}
{\cal F}_1(\tau)&\equiv&\frac 12\left\{e^{-(\Omega_p\tau/2)^2}+
\sqrt{\pi}\;\xi\left[\mbox{Erf}\left(\frac 1{4\xi}-\frac{\Omega_p\tau}
2\right)+
\mbox{Erf}\left(\frac 1{4\xi}+\frac{\Omega_p\tau}2
\right)\right]\right\}
\nonumber\\
{\cal F}_2(\tau)&\equiv&\frac 12\left\{\Lambda(\tau/\tau_\theta)\;
\exp\left[
{-\left(\frac{\Omega_p\tau}2\right)^2\left(\frac{\cos\theta+\sin\theta}
{\cos\theta-\sin\theta}\right)^2}\right]-2\sqrt{\pi}\;\xi\;B(\tau/\tau_\theta)
\mbox{Erf}\left(\frac{1-|\tau|/\tau_\theta}{4\xi}\right)
\right\}
\;\labell{defdeglif},
\end{eqnarray}\end{widetext}
with $\Lambda(x)=1-|x|$ and $B(x)=1$ for $|x|<1$, and
$\Lambda(x)=B(x)=0$ elsewhere. The function ${\cal F}_1$ gives an
envelope for an oscillation at the pump frequency $\omega_p$, while
${\cal F}_2$ is a contribution that disappears both for the $|{\rm
TB}\rangle$ state and for extended-phase-matched crystals, i.e. in the
limit of $\xi\to\infty$. Notice that in this case
Eq.~(\ref{pipimeglio}) reduces to Eq.~(\ref{pipiumz}). In
Fig.~\ref{f:dipsbis} some example plots of coincidence graphs are
given for different values of the phase-matching parameter
$\theta$. As a measure of fringe visibility for the MZ interferometer,
we use the quantity
\begin{eqnarray}
V_{MZ}&\equiv&\frac{P_+(0)-P_+(\pi/\omega_p)}
{P_+(0)+P_+(\pi/\omega_p)}\nonumber\\&=&
\frac{1+[{\cal F}_1(\pi/\omega_p)-{\cal F}_2(\pi/\omega_p)]}
{3-[{\cal F}_1(\pi/\omega_p)-{\cal F}_2(\pi/\omega_p)]}
\;\labell{caleffe},
\end{eqnarray}
which is plotted as a function of the crystal length $L$ for different
values of $\theta$ in Fig.~\ref{f:visib}. In Eq.~(\ref{caleffe}) the
fringe maximum is estimated by $P_+(0)$, and the fringe minimum is
estimated by $P_+(\pi/\omega_p)$; these quantities are easily measured
experimentally. Note that for all interesting cases ${\cal
F}_1(\pi/\omega_p)\geqslant{\cal F}_2(\pi/\omega_p)$, so that
Eq.~(\ref{caleffe}) implies $V_{MZ}\geqslant 33\%$. This lower bound
is reached in the limit of very long crystals $L\to\infty$ if
$\theta\neq -\pi/4$ as shown in Fig.~\ref{f:visib}. Analogous to the
Hong-Ou-Mandel interferometer, the Mach-Zehnder visibility is always
maximum for extended phase-matched crystals ($\theta=-\pi/4$).

\begin{figure}[hbt]
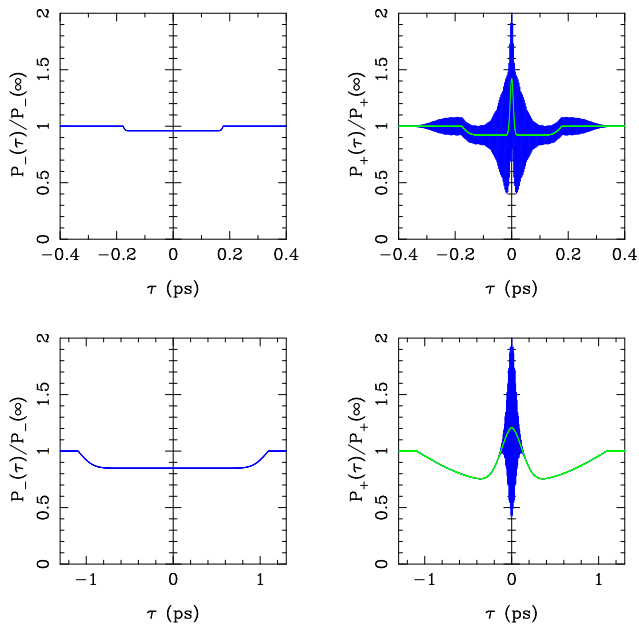

\begin{center}\epsfxsize=.45
\hsize\leavevmode\epsffile{figure6a.eps}
\hspace{.5 cm}\epsfxsize=.45
\hsize\leavevmode\epsffile{figure6b.eps}
\end{center}
\begin{center}\epsfxsize=.45
\hsize\leavevmode\epsffile{figure6c.eps}
\hspace{.5 cm}\epsfxsize=.45
\hsize\leavevmode\epsffile{figure6d.eps}
\end{center}
\caption{Plots of the expected coincidence rates for different
phase-matching conditions. The two left plots refer to the HOM
interferometer setup; the two right plots refer to the MZ
interferometer setup. The parameters in the plots are $\theta=\pi/5$
for the two top plots and $\theta=-\pi/6$ for the two bottom plots;
$\gamma=8\times 10^{-5}$ ps/$\mu$m; $\omega_p=2\times 10^{15}$
s$^{-1}$; $\Omega_p=4\times 10^{13}$ s$^{-1}$; $L=2$ cm. The
coincidence rates for the extended phase-matching case
($\theta=-\pi/4$) were plotted in Fig.~\ref{f:dip}. The degradation of
the HOM dip is evident from the left plots. The presence of the
function ${\cal F}_2$ in Eq.~(\ref{pipimeglio}) is responsible for the
fringe modulation in the plots on the right, as compared with the
$|\Psi_{epm}\rangle$ case shown in Fig.~\ref{f:dip} for which ${\cal
F}_2=0$ prevails.}
\labell{f:dipsbis}\end{figure}

\begin{acknowledgments}
The Authors thank I. A. Walmsley for making them aware of reference
{\cite{walms}}. This work was supported by the DoD Multidisciplinary
University Research Initiative (MURI) program administered by the Army
Research Office under Grant DAAD 19-00-1-0177 and by the National
Reconnaissance Office.
\end{acknowledgments}

\end{document}